\documentclass[12pt]{article}
\usepackage{amstext}
\usepackage{bm}
\usepackage{amssymb,amsmath,mathrsfs,setspace}
\usepackage{graphicx,psfrag,epsf}
\usepackage{array}
\usepackage{natbib}
\usepackage{epsfig}
\usepackage{amssymb}
\usepackage{url}
\usepackage{verbatim}
\usepackage{comment}
\usepackage{color}
\usepackage{graphicx,psfrag,subfigure,amsmath,amssymb,appendix,multirow,float,threeparttable,abstract}
\newtheorem{theorem}{Theorem}[section]
\newtheorem{lemma}[theorem]{Lemma}
\newtheorem{proposition}[theorem]{Proposition}
\newtheorem{corollary}[theorem]{Corollary}

\usepackage[top=1in, bottom=1in, left=1in, right=1in]{geometry}
\fontsize{12}{0}
\numberwithin{equation}{section}
\date{}

\usepackage{comment}

\begin{document}


\newcommand{\blind}{0} 



\if0\blind
{
	\title{\bf Efficient Signal Inclusion With \\
		Genomic Applications \thanks{Address for correspondence: X. Jessie Jeng, Department of Statistics, North Carolina State University, SAS Hall, 2311 Stinson Dr., Raleigh, NC 27695-8203, USA. E-mail: xjjeng@ncsu.edu. 
			}\hspace{.2cm}}
	\author{X. Jessie Jeng$^1$, Teng Zhang$^1$ and Jung-Ying Tzeng$^{1,2,3}$ \\ 
			1. Department of Statistics, North Carolina State University\\
			2. Institute of Epidemiology and Preventive Medicine,\\ 
			National Taiwan University, Taipei, Taiwan\\
			3. Department of Statistics, National Cheng-Kung University, \\
			Tainan, Taiwan
			}
	\maketitle
} \fi

\if1\blind
{
	\bigskip
	\bigskip
	\bigskip
	\begin{center}
		{\LARGE\bf Efficient Signal Inclusion With \\ Genomic Applications \\}
	\end{center}
	\medskip
} \fi

\hspace{-0.5in}


\begin{abstract}
This paper addresses the challenge of efficiently capturing a high proportion of true signals for subsequent data analyses when sample sizes are relatively limited with respect to data dimension. We propose the signal missing rate as a new measure for false negative control to account for the variability of false negative proportion.
Novel data-adaptive procedures are developed to control signal missing rate without incurring many unnecessary false positives under dependence.
We justify the efficiency and adaptivity of the proposed methods via theory and simulation.
The proposed methods are applied to GWAS on human height to effectively remove irrelevant SNPs while retaining a high proportion of relevant SNPs for subsequent polygenic analysis. 
\end{abstract}


\section{Introduction}

High-throughput technology in biology stimulates new challenges in high-dimensional data analysis. For example,  
recent genomic studies have suggested complex, polygenic bases for complex traits --- hundreds of genetic variants are involved in conferring disease risk; individual variants have low effect but variants in aggregate modify disease susceptibility at the gene, pathway or network level. One major goal of the high-throughput biological research, such as genome-wide association studies (GWAS), epigenome-wide association studies (EWAS), and expression quantitative trait locus (eQTL), is to elucidate the joint  mechanisms of a genome-wide set of genomic variables on the trait of interests. 
The exploration of polygenicity requires sophisticated approaches based on simultaneously analyzing  genome-wide variables, e.g., 
pathway based analysis (\cite{Kao2017} and reference therein), polygenic modeling for assessing SNP main or interaction effects 
(\cite{waldmann2013} and reference therein, \cite{wu2010}, \cite{Hung2016}) or for genetic prediction (\cite{abraham2013} and reference therein).
However, due to the high dimension of genome-wide variables and limited sample size, these simultaneous analyses have to be coupled with a pre-screening step to reduce the data dimension.

Pre-screening can greatly impact subsequent analyses. When individual variants have small effect, pre-screening that is too stringent may fail to capture them for follow-up studies. On the other hand, a pre-screening that is too liberal can hurt the performance of the subsequent simultaneous analyses by including too many noise variables. 
In current practice, pre-screening on genome-wide variables is often performed by selecting the SNPs with p-values less than an arbitrary threshold (e.g., p-value $< 0.001$ in \cite{zhou2011}, and $<0.0001$ in \cite{wu2009}).
In this work, we aim to develop a data-driven method that is adaptive to the underlying data features so that a high proportion of signals can be selected without incurring many unnecessary noise variables.

One major challenge in developing a data-adaptive method for signal inclusion is how to effectively accommodate the unknown signal information such as signal sparsity and intensity. When signals are much rarer than noise, inference based on signal information  is more challenging than inference based on noise distribution. Consequently, retaining signals through false negative control requires different techniques from those used for false positive control. 
Another challenge for data-adaptive signal inclusion is how to accommodate  dependence among variables in real applications. For example, in genomic data analysis, a Single Nucleotide Polymorphism (SNP) is usually strongly correlated with the SNPs nearby due to linkage disequilibrium. The dependence among SNPs can dramatically effect the test statistics and confound inference.

In this paper, we propose a new analytic framework for efficient signal inclusion under dependence. 
We first discuss sensible criteria for false negative control and propose a new measure called signal missing rate (SMR). 
Compared to existing measures, SMR assesses the exceedance probability of false negative proportion and incorporates the variability of  false negative proportion into inferences. 

Next, we develop data-adaptive procedures to control SMR under dependence. The first procedure, conservative SMR (cvSMR), utilizes existing techniques in multiple testing to control false discovery proportion at a stochastic level involving signal information and, consequently, control the measure of SMR at a low level. 
cvSMR is quite intuitive and easy to implement. However, it tends to be overly conservative for false negative control and can include too many noise variables.

In order to improve the efficiency of signal inclusion,  we propose the second method, Adaptive SMR (AdSMR). The main difference between AdSMR and cvSMR is that AdSMR implements a much relaxed critical sequence in its selection rule, which results in a smaller subset of selected variables. 
The new critical sequence is established by novel theoretical analysis on the variability of false negative proportion  through concentration properties of order statistics under  dependence. 
The improved AdSMR procedure and the 
new analytic techniques guarantee the control of SMR at a degenerating level and the control of unnecessary false positives.    
Although the implementation of AdSMR does not need signal information, the cut-off position of AdSMR automatically vary with signal sparsity and intensity, and, as a result, when signal intensity become stronger, both false negative and false positive can be better controlled by  AdSMR.


A by-product in the study is a consistent estimator for the number of signals under block dependence that is widely observed in genomic data. Existing studies on signal proportion estimation mainly assume independence \citep{meinshausen2006estimating, jin2007estimating}. Consistent estimation under dependence is not only useful for signal inclusion as described in this paper, but also valuable in other areas, such as to improve the performance of FDR-based methods in multiple testing.

We compare the finite-sample performances of the proposed AdSMR method and existing methods in simulation. The simulation settings include different sparsity and intensity levels of signals, block dependence with various block sizes, sparse dependence without block structures, and dependence structure from a multi-factor model. While all the methods seem to be effective in false negative control, AdSMR generally outperforms other methods in incurring less false positives.

We apply AdSMR to a GWA analysis on human height using the CoLaus data, in which all  $340,359$ autosomal SNPs explain $53.7\%$ of the phenotypic variability of human height. Multiple testing based on the full set of SNPs identifies zero significant candidates because individual variants have small effects. In order to significantly reduce data dimension and carry as many relevant SNPs to subsequent polygenic analyses, we apply AdSMR and select only a small proportion ($0.021$) of the total SNPs, which explains nearly all the $53.7\%$ of the height variation attained using the full set of SNPs. 
We further apply penalized regression on the SNPs selected by AdSMR and narrow down the number of selected SNPs to $1,563$. The estimated heritability is still close to $53.7\%$ based on only the $1,563$ SNPs.
The selected subset would include a high proportion of truly relevant SNPs for further  downstream analyses such as gene annotation, pathway mapping, polygenic risk score, etc.

The rest of the paper is organized as follows. Section \ref{sec:method_theory} introduces the signal missing rate and develop procedures for SMR control under dependence. Consistent estimation of signal proportion is also discussed. Section \ref{sec:simulation} demonstrates the finite-sample performance of the proposed methods in simulation. Comparisons with other methods are provided. Section \ref{sec:realdata} presents an application of our method to genomic data analysis. Concluding remarks are provided in Section 5.

\section{Efficient Signal Inclusion Under Dependence} \label{sec:method_theory}

\vspace{-0.1in}
\subsection{Signal Missing Rate} \label{sec:FormulationSMR}

We first discuss sensible measures for false negatives. 
Table  \ref{tab:classification} summarizes notations in classification of variables where TP, FN, FP, and TN are numbers of true positives, false negatives, false positives, and true negatives, respectively;  $s$ is the total number of signal variables. Our goal of signal inclusion corresponds to seeking low $FN/s$.

\begin{table}[h]
	\begin{center}
		\caption{Classifications of variables.}\label{tab:classification}
		\begin{tabular}{llll}\hline
			\textbf{} & \textbf{Selected} & \textbf{Not selected}  & \textbf{Total}  \\
			\hline
			\textbf{Signal} &   {TP}   &  {FN} &  $s$\\
			\textbf{Noise} &    {FP} &  {TN} &  $m-s$ \\
			& $R$ & $m-R$ & $m$ \\
			\hline
		\end{tabular}
	\end{center}
	\vspace{-0.1in}
\end{table}

In the multiple testing literature, False Nondiscovery Rate (FNR) has been proposed as an analogue of False Discovery Rate (FDR) to measure false negatives.  It is defined as the expectation of the proportion of false negatives among the unselected variables, $E(\text{FN}/(m-R))$ \citep{genovese2004stochastic, Sarkar2006}. 
The notion of FNR, unfortunately, does not suit our need for signal inclusion, because FNR is mostly very close to zero when signals are sparse, and a large (or small) FNR does not correspond to a high (or low) $FN/s$. 
We provide a simulation example to illustrate this point in Supplementary Material.

Recently, \cite{cai2016optimal} constructed the Missed Discovery Rate (MDR) for false negative control.  MDR is the expected value of false negative proportion, $E(FN/s)$.
To utilize MDR for false negative control, all the variables are ranked by their estimated local FDR values ($\widehat{Lfdr}$). 
Because $E(\text{FN})$ can be approximated by $\sum(1-\widehat{Lfdr})$, given an estimate for $s$, a cut-off position on the ranked local FDR values can be determined to control MDR at a pre-fixed level.
MDR control is intuitive and easy to implement. 
However, the measure of MDR does not consider the variability of false negative proportion; and the control of MDR   has not been studied under dependence. 

In this paper, we propose a new measure for false negative control called Signal Missing Rate (SMR). SMR is defined as 
\vspace{-0.15in}
\begin{equation} \label{def:SMR}
SMR^{\epsilon}=P\left(FN/s > \epsilon\right),
\vspace{-0.1in}
\end{equation}
where $\epsilon>0$ is a constant between $0$ and $1$. Signal missing rate evaluates the probability of neglecting at least a certain proportion of signals.  By controlling SMR at a low level with a small $\epsilon$, a high proportion of signals can be captured. 
Compared to MDR, SMR measures the exceedance probability of false negative proportion and incorporates the variability of $FN/s$ into inference.


\vspace{-0.1in}
\subsection{Controlling SMR under dependence}
 
To assure generality of our work,  we do not assume any specific distribution for the test statistic. 
Specifically, define $I_0$ and $I_1$ as collections of indices of the noise and signal variables, respectively. Let $P_j$ be the $p$-value of the $j$th variable. 
Assume 
\begin{equation} \label{def:p_model}
P_j \sim U\cdot 1\{j \in I_0\} + G \cdot 1\{j \in I_1\}, \qquad j = 1, \ldots, m,
\end{equation}
where $U$ represents the cumulative distribution function of the uniform distribution at [0,1] and $G$ is some unknown cumulative distribution function dominating $U$, i.e., $G(t) > U(t)$ for all $t \in (0, 1)$. 
This mixture model on $p$-values in (\ref{def:p_model}) provides a convenient framework for large-scale inference. It can be used in a wide range of applications as long as the baseline distribution of the noise can be reasonably estimated from either asymptotic or empirical approaches, such as by permutation or parametric bootstrap.

We define the signal missing rate of a procedure selecting the top $k$ candidates along the ranked $p$-values as
\[
SMR^{\epsilon}(k) = P(FN(k) / s > \epsilon),
\]
where $\epsilon\in (0, 1)$ is a constant and $FN(k)$ represents the number of false negatives for selecting the top $k$ candidates, which equals to the number of true signals ranked after $k$.  
 
We develop two procedures for SMR control. Both procedures are easy to implement in applications. 
The first procedure is more in line with existing techniques in multiple testing.
The second procedure improves the efficiency of the first approach by developing new analytic techniques based on concentration inequalities of order statistics under dependence.

\subsubsection{The conservative SMR procedure} 

Suppose that we know the number of signals $s$, then a procedure controlling $FDP (= FP/R)$ at the level of $(R-s)/R$ includes the number of signals as $TP = R-FP = R-FDP\times R \ge s$. 
Therefore, one can modify a method that controls FDP at the level of $(R-s)/R$ to include a high proportion of signals. Motivated by this idea, we develop the conservative SMR (cvSMR), a procedure that determines the cut-off position on the ranked $p$-values $p_{(1)}, \ldots, p_{(m)}$ by 
\begin{equation} \label{cut_strong}
{k}^*_{cv}  = \hat s + \min \{j \ge 1: p_{(\hat s+j)}\leq {j \over m} \alpha\}1\{\hat{s}>t_1\},
\end{equation}
where $\hat s$ is an estimate for the number of signals, $\alpha$ is a prefixed small constant, and $t_1=\max\{j: p_{(j)} < \alpha_m/m\}$ with $\alpha_m=o(1)$. 
The top $\{1, \ldots, {k}^*_{cv} \}$ candidates are selected.

cvSMR is a step-down procedure with critical sequence $\alpha_j = (j / m)\alpha$ that is frequently used in methods controlling FDR or FDP.
Compared to an existing step-down procedure studied in \cite{Lehmann2005}, cvSMR uses an opposite sign when comparing $p$-value with the critical sequence. This is because the procedure starts at the position $\hat s$, where FDP is not controlled in general, and stops as soon as FDP is controlled at a desirable level. On the other hand, the procedure in \cite{Lehmann2005} starts from the first position where FDP is controlled and stops once FDP cannot be controlled. 
cvSMR also differs in the index of $\alpha_j$, where $j$ is not the index of the ordered $p$-values (k), but $k-\hat s$.
It can be proved that the step-down procedure of cvSMR controls FDP at a stochastic level of $({k}^*_{cv} -\hat s)/{k}^*_{cv} $. Consequently, the number of true signals included in the top ${k}^*_{cv}$ variables is greater than $\hat s$ with high probability. 
 
The following theoretical results show that  cvSMR asymptotically controls SMR at the level of $\alpha$. 
$P_1^0, \ldots, P_{m-s}^0$ denote the $p$-values corresponding to the $m-s$ noise variables and $P_1^1, \ldots, P_{s}^1$ denote the $p$-values corresponding to the $s$ signals. We consider the same dependence condition as in \cite{Lehmann2005}:  for any $j = 1, \ldots, m-s$, 
\begin{equation} \label{cond:jointP_1}
P(P_j^0 \le u | P^1_1, \ldots, P^1_s) \le u.
\end{equation}
This condition says that the $p$-value of a noise variable is conditionally dominated by a uniform distribution. This condition allows arbitrary joint dependence within noise variables and within signal variables.

\begin{proposition}\label{prop:control_1}
	Consider model (\ref{def:p_model}) under condition (\ref{cond:jointP_1}). Given a consistent estimator $\hat s$ for the number of signals and a constant $\alpha$ for the SMR control level, cvSMR  asymptotically controls $SMR^\epsilon$ at the level of $\alpha$ for any $\epsilon>0$, i.e., 
	\begin{equation} \label{eq:control_1}
	SMR^{\epsilon}({k}^*_{cv}) \le \alpha + \Delta_m,
	\end{equation}
	where  $\Delta_m =o(1)$ for any $\epsilon>0$. 
\end{proposition}    
Note that $\Delta_m =o(1)$ as long as $\epsilon$ is a constant. If $\epsilon = \epsilon_m \to 0$, the asymptotic control on SMR by cvSMR may not hold. This follows our
intuition that the control of false negatives is harder when smaller number of false negatives is allowed.

Generally speaking, cvSMR would work well in situations where FDR/FDP methods work well. 
However, cvSMR inherits the same issue as FDR/FDP methods and tends to be conservative under high dimensionality. Here, the conservativeness is on false negative control, which means that too many variables could be selected. In fact, this disadvantage can be more severe for cvSMR as the event $\{p_{(\hat s+j)} \le (j/m) \alpha\}$ is less likely to happen than the event $\{p_{(j)} \le (j/m) \alpha\}$ in FDR/FDP methods. Table 2 presents the number of selected candidates in a simulation example with $m=5000$ and $s = 250$. The test statistics are generated from multivariate normal distribution $N(\mathbf{A}, \mathbf{\Sigma})$, where $A_j = 0$ for noise and $A_j = \mu >0$ for signal.
The covariance matrix $\mathbf{\Sigma}$ is a block-diagonal matrix with equal block size $l=50$ and within-block correlation $\rho=0.7$. cvSMR with $\alpha=0.1$ appears too conservative in this example by selecting almost all variables.

\begin{table}[h]
	\centering
	\caption{The average cut-off positions from $100$ replications with $m=5000$ and $s=250$.}
	\begin{tabular}{|l|c c c|}
		\hline
		 &  $\mu=3.5$ & $\mu=4.5$ & $\mu=5.5$ \\
		\hline
		 cvSMR &  5000 & 5000 & 4905 \\
		 AdSMR &  348 & 300 & 288 \\
		\hline
	\end{tabular}
	\label{tab:cutCompare}
\end{table}

Because cvSMR controls $SMR^\epsilon$ for arbitrarily small constant  $\epsilon>0$,  one way to mitigate the conservativeness of cvSMR is to weaken the control of false negatives and  allow for a fixed proportion of false negatives. 
Consequently, the critical sequence $\alpha_j = (j/m) \alpha$ can be relaxed by involving the fixed proportion. 
Considering our motivation for retaining as many true signals as possible for subsequent data analysis, we would like to propose a different strategy to significantly reduce the number of false positives without weakening the theoretical control on false negatives.

\subsubsection{The Adaptive SMR Procedure} 
In order to develop a method that incurs less false positives and is more applicable in Big Data applications, we propose the second procedure, the Adaptive SMR (AdSMR). 
AdSMR has the cut-off position on the ranked $p$-values as
\begin{equation} \label{cut_weak}
{k}^* = \hat s + \min \{j \ge 1: p_{(\hat s+j)}\leq {b_j}\}1\{\hat{s}>t_1\}.
\vspace{-0.1in}
\end{equation}
Similar to (\ref{cut_strong}), $\hat s$ is an estimate of the number of signals and $t_1=\max\{j: p_{(j)} < \alpha_m/m\}$ with $\alpha_m=o(1)$. The key difference between (\ref{cut_strong}) and (\ref{cut_weak}) is that AdSMR has the critical sequence $b_j$ defined as the median of $Beta(j, m-\hat s -j+1)$. 
The rationale to use beta distribution to determine $b_j$ is because our proposed method is based on ranked $p$-values, and the $j$-th ranked $p$-value of $m-\hat s$ noise variables follows  $Beta(j, m-\hat s-j+1)$ under independence. Therefore, it is natural to utilize  $Beta(j, m-\hat s-j+1)$ to perform inference.

Although the median of a beta distribution does not have an explicit form, it is known that the median is bounded by the mode and mean of beta distribution. Therefore, 
\begin{equation} \label{dj_bound}
{j-1 \over m- \hat s -1} < b_j < {j \over m-\hat s +1},
\end{equation} 
and $b_j$ is approximately $\alpha^{-1}$ times as large as $\alpha_j$ of cvSMR. Larger $b_j$ results in less variables being selected as shown in Table \ref{tab:cutCompare}.

\begin{figure}[H]
	\centering
	\caption{A sequence of ordered $p$-values. $T_1$ denotes the location right before the first noise variable and $T_2$ denotes the location of the last signal variable.}
	\includegraphics[scale = 0.5]{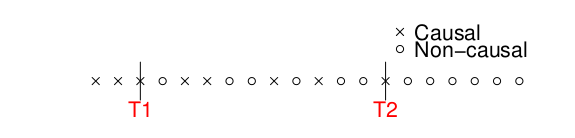}
	\label{t1t2}
	\vspace{-0.2in}
\end{figure}

To justify this new procedure in theory, existing techniques for FDP control cannot be used anymore. We develop novel techniques to analyze the procedure based on concentration properties of the ordered $p$-values under dependence.
Figure \ref{t1t2} illustrates a sequence of ordered $p$-values, where $T_1$ denotes the location before the first noise variable and $T_2$ denotes the location of the last signal variable. Signals and noise are mixed indistinguishably between $T_1$ and $T_2$.
We show that the proposed AdSMR method is able to capture a high proportion of signals ranked before $T_2$ and, at the same time, avoid unnecessary false positives ranked after $T_2$. 

Let $L$ be the total number of noise variables ranked before $T_2$, then $L = T_2-s$.  
Note that $L$ is a random variable varying from sample to sample. Generally speaking, lower signal intensity results in larger $L$. The specific relationship depends on the model that generates the data.   
In this section, we assume that $L$ is bounded almost surely by a number $\bar{l}$, and
\begin{equation} \label{cond:barL}
1 \ll \bar{l} \log(\bar{l}) \ll \min(s^2, \sqrt{ms}).
\end{equation}   
Condition (\ref{cond:barL}) says that the number of indistinguishable noise is not too large. For example, in an association study with $500,000$ total variables and $100$ truly associated signals, we request $\bar{l} \log(\bar{l}) \ll 7,071$. This condition is fairly general as it allows the existence of weak signals that may rank after many noise variables and ``pseudo" signals which are indeed noise but show up as signals due to high dependence with true signals. We note that the number of such ``pseudo" signals is often much smaller than the total number of variables in applications when true signals are sparse.

Our next condition is on dependence in the data. 
Let $P_{(1)}^0, \ldots, P_{(m-s)}^0$ denote the ordered $p$-values corresponding to the $m-s$ noise variables. Assume that, for any $r = 1, \ldots, \bar{l}$, 
\begin{equation} \label{cond:jointP_2}
P(P_{(r)}^0 \le u | P^1_{1}, \ldots, P^1_{s}) \le c_1 F_{r}(u),
\end{equation}
where $F_{r}(\cdot)$ is the conterpart of the left side probability for independent $p$-values, and $c_1\ge 1$ is some constant.  
Since order statistics of independent noise $p$-values follow beta distribution,  $F_{r}(\cdot)$ is the cumulative distribution function of $Beta(\nu_1, \nu_2)$ with $\nu_1=r$ and $\nu_2 = m-s -r+1$. This condition essentially says that the $\bar{l}$ smallest noise $p$-values (given the signal $p$-values) can be more or less extreme than their counterparts under independence as long as $c_1$ is a constant no less than 1. 
There are no constraints on the noise $p$-values ranked after $\bar{l}$ or the $p$-values of signal variables. 

The following theorem  shows that  given the above conditions, AdSMR has a degenerating SMR, which is equivalent to say that the FN proportion/sensitivity of AdSMR converges to $0$/$1$ in probability. 
\begin{theorem}\label{thm:control_2}
	Consider model (\ref{def:p_model}) under conditions (\ref{cond:barL}) and (\ref{cond:jointP_2}). Given a consistent estimator $\hat s$ for the number of signals, AdSMR has a degenerating $SMR^\epsilon$ for any $\epsilon>0$, i.e.,
	\begin{equation} \label{screening_2}
	SMR^{\epsilon}(k^*) \to 0
	\end{equation}
	as $m \to \infty$ for any constant $\epsilon>0$.
\end{theorem}      
Comparing Theorem \ref{thm:control_2} with Proposition \ref{prop:control_1}, it can been seen that  AdSMR and 
cvSMR control SMR differently. cvSMR asymptotically controls SMR at a prefixed level $\alpha$, whereas AdSMR controls at a degenerating level.  
The theoretical justification coupled with the more relaxed critical sequence make AdSMR a more efficient method for false negative control. 

The asymptotic result in Theorem \ref{thm:control_2} holds for any constant $\epsilon>0$ but may not hold for $\epsilon \to 0$. In other words, AdSMR may allow a number of false negatives as long as the number is not greater than a proportion of the total number of signals.

Next, we show that AdSMR can avoid selecting unnecessary false positives. Recall the locations of $T_1$ and $T_2$ in Figure \ref{t1t2}. Although signal and noise variables mix indistinguishably between $T_1$ and $T_2$, noise variables ranked after $T_2$ should be avoided. The next theorem shows that under suitable conditions on the dependence and the estimator $\hat s$, AdSMR controls the selection of noise variables ranked after $T_2$.    
\begin{theorem}\label{thm:FP}
	Consider model (\ref{def:p_model}). Define $T_2$ and $L$ as in Figure \ref{t1t2}. Assume that $L \gg 1$ with probability tending to 1, and for some small constant $\delta>0$,
	\begin{equation} \label{cond:FPcontrol}
	P(P_{(L+k)}^0 > u_k,  k = 1, \ldots, \delta L | P^1_{1}, \ldots, P^1_{s}) \le  c_2 H(u_k, k = 1, \ldots, \delta L), 
	\end{equation}
	 where $H(\cdot)$ is the counterpart of the left side probability for independent $p$-values, and $c_2\ge 1$ is some constant. Then, AdSMR with $\hat s$ satisfying $P(\hat s < s) \to 1$ has 
	\begin{equation} \label{eq:FP}
	P(k^* > (1+\delta) T_2) \to 0.
	\end{equation}
\end{theorem}
Condition (\ref{cond:FPcontrol}) is quite general as $c_2$ is an arbitrary constant no less than 1. 
Theorem \ref{thm:control_2} and \ref{thm:FP} imply that AdSMR achieves both SMR control and false positive control when (a) dependence conditions in (\ref{cond:jointP_2}) and (\ref{cond:FPcontrol}) are satisfied, (b) signal intensity is strong enough so that condition (\ref{cond:barL}) is satisfied, and (c) $P((1-\delta) < \hat s < s) \to 1$ for arbitrarily small constant $\delta>0$. The $\hat s$ estimator studied in the next section has the property in (c).

\vspace{-0.1in}
\subsection{AdSMR with MR estimator} \label{sec:algorithm}

An important component of the SMR control-based methods is a consistent estimator for the number of signals.  
Estimation of signal proportion among all variables ($\pi = s/m$) has inspired profound research in high-dimensional inference. 
For example, \cite{storey2002direct}, \cite{genovese2004stochastic}, and  \cite{jin2007estimating} have developed consistent estimators for relatively dense signals with proportion $\gg m^{-1/2}$;
\cite{cai2007estimation} has considered sparse signals with proportion $\le m^{-1/2}$; and
\cite{meinshausen2006estimating} has considered both dense and sparse signals for proportion estimation.
However, existing studies mainly focus on independent variables.  Rigorous analysis under realistic dependence structure is scarce.

\begin{figure}[h!]
	\centering
	\caption{Heatmap of the absolute value of correlations for $50$ SNPs in CoLaus data.}
	\includegraphics[scale = 0.5]{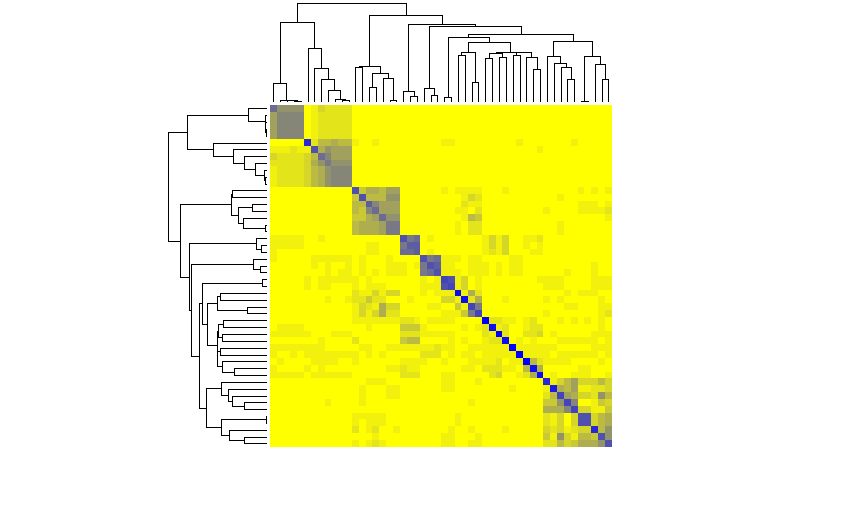}
	\label{heatmap1}
	\vspace{-0.3in}
\end{figure}

In genomic data analysis, the covariance matrix of $p$-values of SNPs often exhibits a block structure. In Figure \ref{heatmap1}, we present heatmap of the absolute values of sample correlations of the first $50$ SNPs in Chromosome 1 from Cohorte Lausannoise (CoLaus) study samples. Details of the real data are described in Section \ref{sec:realdata}. The heatmap shows blocks of high correlations along the diagonal regions with different block sizes. Such dependence structures are frequently observed in genomic data
\citep{Efron07, fan2012estimating}. 

In this section, we study the consistency of the estimator developed in \cite{meinshausen2006estimating} in the situation where variables have block dependence. 
We refer to this estimator as the MR estimator which 
is constructed as follows. 
Define the empirical distribution of $p$-values 
\begin{equation}\label{pempirical}
F_m(t) = \frac{1}{m}\sum_{j=1}^m 1\{P_j \leq t\} = (1 - \pi) U_{m_0}(t) + \pi G_{s}(t),
\end{equation}
where $U_{m_0}$ and $G_{s}$ denote the empirical distributions of the $p$-values corresponding to $m_0 (= m-s)$ noise variables and $s$ signals, respectively. Similarly, denote $U_{m}$ as the empirical distributions of the $p$-values when all $m$ candidates are noise. Define
\begin{equation} \label{Vm}
V_m = \sup_{t \in (0,1)} \frac{U_m(t)-t}{\sqrt{t(1-t)}},
\end{equation}
and denote $c_m$ as a bounding sequence of $V_m$ such that  $mc_m$ is monotonically increasing with $m$ and $P(V_m > c_m) = \alpha_m \to 0$.
Then, the MR estimator for $ \pi$ is constructed as
\begin{equation} \label{estimator}
\hat{\pi}_{MR} = \sup_{t \in (0, 1)} \frac{F_m(t)-t- c_m \sqrt{t(1-t)}}{1-t}.
\end{equation}

Consistency of $\hat{\pi}_{MR}$ has been shown for independent $p$-values in \cite{meinshausen2006estimating}. 
In this section, we study the consistency of $\hat \pi_{MR}$ under block  dependence and implement $\hat \pi_{MR}$ into AdSMR. 
Assume that $\{P_j\}_{j=1}^m$ can be divided into independent groups with arbitrary dependence in each group. 
Denote $l$ as the upper bound of the group sizes and assume
\vspace{-0.1in}
\begin{equation} \label{def:blocksize}
l=O(m^{\kappa}) \quad \text{for some constant } \kappa \in [0,1).
\vspace{-0.1in}
\end{equation}
The next theorem summarizes the conditions for $\hat \pi_{MR}$ consistency and the SMR control of AdSMR with MR estimator.
\begin{theorem} \label{thm:unify}
	Consider model (\ref{def:p_model}) under condition (\ref{cond:jointP_2}) and block dependence in (\ref{def:blocksize}). Let $\pi = m^{-\eta}$ for some $\eta\in [0,1)$. Assume either one of the following conditions: \\
	(i) $\eta \in [0, (1-\kappa)/2)$, $\inf_{t \in (0,1)} G'(t) = 0$, and $1 \ll \bar{l} \log(\bar{l}) \ll m^{1-\eta/2}$. \\
	(ii) $\eta \in [(1-\kappa)/2, 2/3)$, $G(m^{-\tau}) \to 1$ for some $\tau> 2 \eta - (1-\kappa)$, and $1 \ll \bar{l} \log(\bar{l}) \ll m^{1-\eta/2}$. \\
	(iii) $\eta \in [2/3, 1)$, $G(m^{-\tau}) \to 1$ for some $\tau> 2 \eta - (1-\kappa)$, and $1 \ll \bar{l} \log(\bar{l}) \ll m^{2(1-\eta)}$. \\	
	Then $P(1-\delta < \hat{\pi}_{MR} / \pi < 1) \rightarrow 1$ as $m \to \infty$ for arbitrarily small constant $\delta > 0$ and 	
	AdSMR with $\hat s = m \hat \pi_{MR} $ has $SMR^{\epsilon}(k^*) \to 0 $ as $m \to \infty$ for any constant $\epsilon>0$.
\end{theorem}
Theorem \ref{thm:unify} considers the scenario where the same dataset is used to derive $\hat \pi_{MR}$ and to retain signals by AdSMR.  Therefore, the SMR control of AdSMR  would be affected by the conditions for the consistency of $\hat \pi_{MR}$.
Given the general block structure of dependence, condition (\ref{cond:jointP_2})  implies constraints on the block size and within-block dependence. Under the constraints of dependence, conditions in (i) - (iii) can be satisfied if signal intensity is strong enough in different ranges of sparsity level. 

It can be seen that the more sparse the signals (larger $\eta$), the stronger the condition on signal intensity. When $\eta \in [0, (1-\kappa)/2)$, the condition $\inf_{t \in (0,1)} G'(t) = 0$ is quite general and has been described as the ``pure" case in \cite{genovese2004stochastic}. When $\eta \in [(1-\kappa)/2, 1)$, the stronger condition $G(m^{-\tau}) \to 1$ says that the distribution of signal $p$-values is highly concentrated around $0$. One can also see the effect of block size demonstrated through $\kappa$. Generally speaking, the larger the $\kappa$, the stronger the conditions in (i) - (iii).

AdSMR with MR estimator also controls unnecessary false positives ranked after $T_2$ as presented in the following corollary. The proof  duplicates parts of the proofs for Theorem \ref{thm:FP} and \ref{thm:unify}, and thus are omitted.  
\begin{corollary} \label{cor:FP}
	Assume conditions in Theorem \ref{thm:FP} and block dependence in (\ref{def:blocksize}).  AdSMR with $\hat s = m \hat \pi_{MR}$ has 
	$P(k^* > (1+\delta) T_2) \to 0$ for arbitrarily small constant $\delta>0$.
\end{corollary}

The next corollary shows that AdSMR with MR estimator has asymptotically zero false positives in two special scenarios.

\begin{corollary} \label{cor:typeI}
	Assume model (\ref{def:p_model}) with block dependence in (\ref{def:blocksize}). Consider two scenarios: (i) there is no signals exists, i.e. $s=0$; and  (ii) signals are strong enough such that $s \bar G(m^{-\tau_0}) = o(1)$ with $\bar G = 1-G$ and $\tau_0>1$. In both scenarios, AdSMR with $\hat s = m \hat \pi_{MR}$ has $P(FP(k^*) > 0) = o(1)$. 
\end{corollary}

We conclude this section by a complete algorithm for AdSMR with the MR estimator. For simplicity, the same $\alpha_m$ is used to simulate the bounding sequence $c_m$ and to obtain $t_1$ in AdSMR. More specifically, one can simulation $V_m$ from the empirical null distribution. Then, $c_m$ can be determined as the $(1-\alpha_m)$th quantile of the empirical distribution of $V_m$ from 1000 simulations. 
To save computation, we approximate $b_j$ by $j/(m-\hat s)$ as shown in (\ref{dj_bound}) and set an upper limit for $k^*$ at $\lfloor m/2 \rfloor$.
The step-by-step algorithm is as follows. A toy example demonstrating the algorithm is provided in Appendix \ref{sec:example}.

\vspace{0.1in}

\noindent \underline{\bf Algorithm 1: AdSMR with MR estimator} 
\vspace{-0.1in}
\begin{enumerate}
	\item Simulate the bounding sequence $c_m$ from the empirical null distribution of $V_m$ with $\alpha_m = 1/\sqrt{\log{m}}$.
	\vspace{-0.1in}
	\item Obtain $\hat \pi_{MR}$ by (\ref{estimator}) using the bounding sequence $c_m$.
	\vspace{-0.1in}
	\item Sort the observed $p$-values as $p_{(1)} \le p_{(2)} \le \ldots\le p_{(m)}$.
	\vspace{-0.1in}
	\item Calculate the cut-off position by $k^*$ in (\ref{cut_weak}) with $\hat s = m \hat \pi_{MR}$, $\alpha_m = 1/\sqrt{\log{m}}$ and $b_j = j /(m-\hat s)$.  Set an upper limit for $k^*$ at $\lfloor m/2 \rfloor$. 
	\vspace{-0.1in}
	\item Select the top ranked candidates with $p$-values $p_{(1)}, \ldots, p_{(k^*)}$.
\end{enumerate}

\vspace{-0.2in}
\section{Simulation Study} \label{sec:simulation}
We compare the finite-sample performances of AdSMR and existing  methods for false negative control. 
These methods include the MDR procedure \citep{cai2016optimal} and BH-FDR with high nominal levels.
For fair comparison, both AdSMR and MDR use $\hat \pi_{MR}$ for proportion estimation. 
MDR aims to control the expectation of FN proportion through estimating local FDR. 
We use the software ``locfdr" to estimate local FDR and apply MDR at the recommended level $1/\log(m)$. BH-FDR with high nominal levels are ad-hoc procedures that apply the original FDR method in \cite{benjamini1995controlling} with high nominal levels 0.5 and 0.7 to capture more signals. 

We demonstrate FN control of these methods by reporting their  FN proportions ($FN/s$). Then, we show their efficiency by reporting the false discovery proportion ($FDP = FP/R$) of these methods. Higher FDP can be viewed as higher price paid  to achieve low FN proportion. 
In addition, we employ the F-measure as a summary metric for FN proportion and FDP \citep{F-measure2011}.  By definition, F-measure is the harmonic mean of precision ($1-FP/R$) and recall ($1-FN/s$) and calculated by $ 2 \times precision \times recall / (precision + recall)$.

We simulate test statistics from multivariate normal distribution $N(\mathbf{A}, \mathbf{\Sigma})$, where $A_j = 0$ for noise and $A_j = \mu >0$ for signal. The locations of the signals are selected randomly.  We consider settings with different signal sparsity and intensity levels and various dependence.

\begin{figure} [h!]
	\centering
	\caption{Comparison of methods in false negative proportion, false discovery proportion, and F-measure for block dependence with $l=50$. The top row has $\pi = 0.1$, and the bottom row has  $\pi = 0.02$.} \label{Fig:smallL}
	\includegraphics[width=6in,height=2.2in]{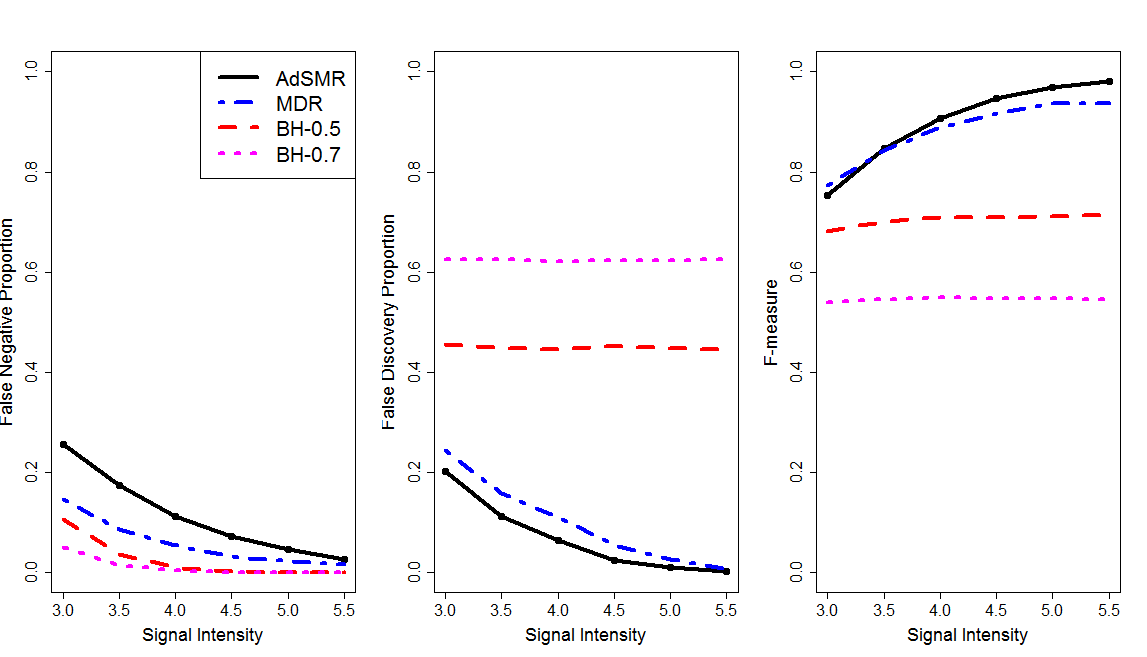}
	\includegraphics[width=6in,height=2.2in]{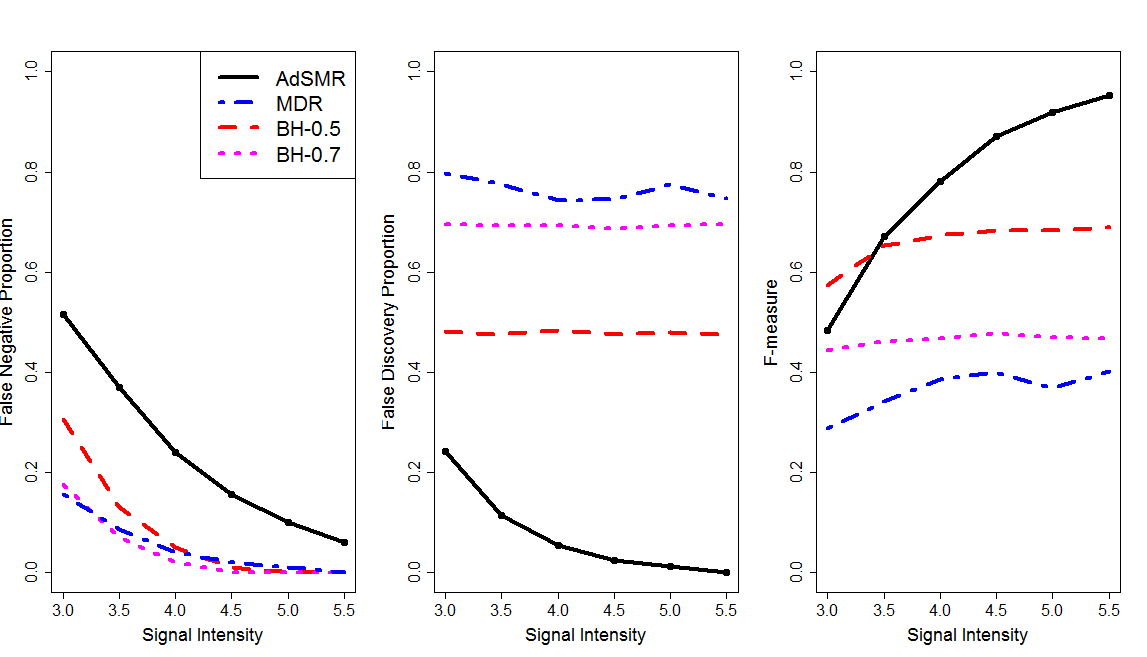}
\end{figure}

Example 1 has data dimension $m=5000$. The covariance matrix $\bf{\Sigma}$ is a block diagonal matrix with equal block size $l=50$ and within-block correlation $\rho = 0.7$. The diagonal elements of  $\bf{\Sigma}$  are set to be 1. We demonstrate different signal sparsity levels: $\pi = 0.02 $ and $0.1$. Signal intensity $\mu$ increases from 3 to 5.5.
Figure  \ref{Fig:smallL} presents the median values of FN proportion, FDP, and the F-measure for all the methods from 100 simulations. It shows that the FN proportion of AdSMR is lower for signals with $\pi=0.1$ than for signals with $\pi=0.02$. This agrees with the theoretical insights of Theorem \ref{thm:unify} where stronger condition on signal intensity is required for more sparse signals. Compared to other methods, AdSMR has slightly larger FN proportion but smaller FDP. AdSMR generally outperforms other methods in F-measure especially for sparse signals with $\pi = 0.02$. 
We follow the convention in high-dimensional screening literatures, e.g., \cite{fan2008sure}, and report the median of these measures. 
Examples of the mean and standard deviation of the measures are provided in Supplementary Material. 

\begin{figure} [h!]
	\centering
	\caption{Comparison of methods in false negative proportion, false discovery proportion, and F-measure for block dependence with $l=200$. The top row has $\pi = 0.1$, and the bottom row has $\pi = 0.02$. } \label{Fig:largeL}
	\includegraphics[width=6in,height=2.2in]{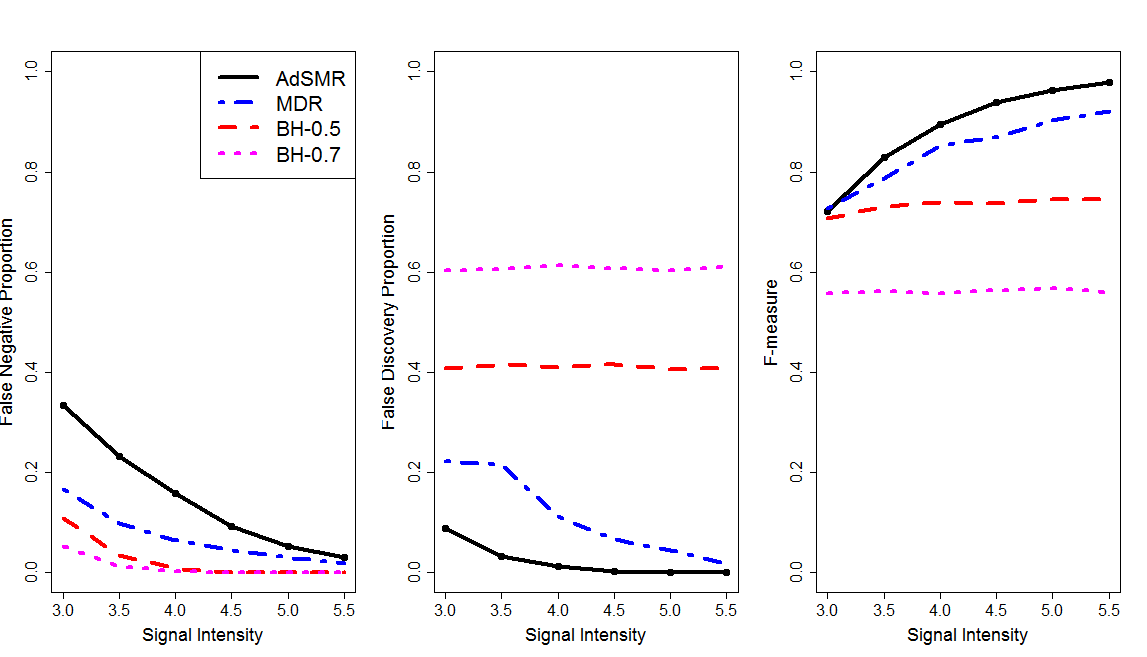}
	\includegraphics[width=6in,height=2.2in]{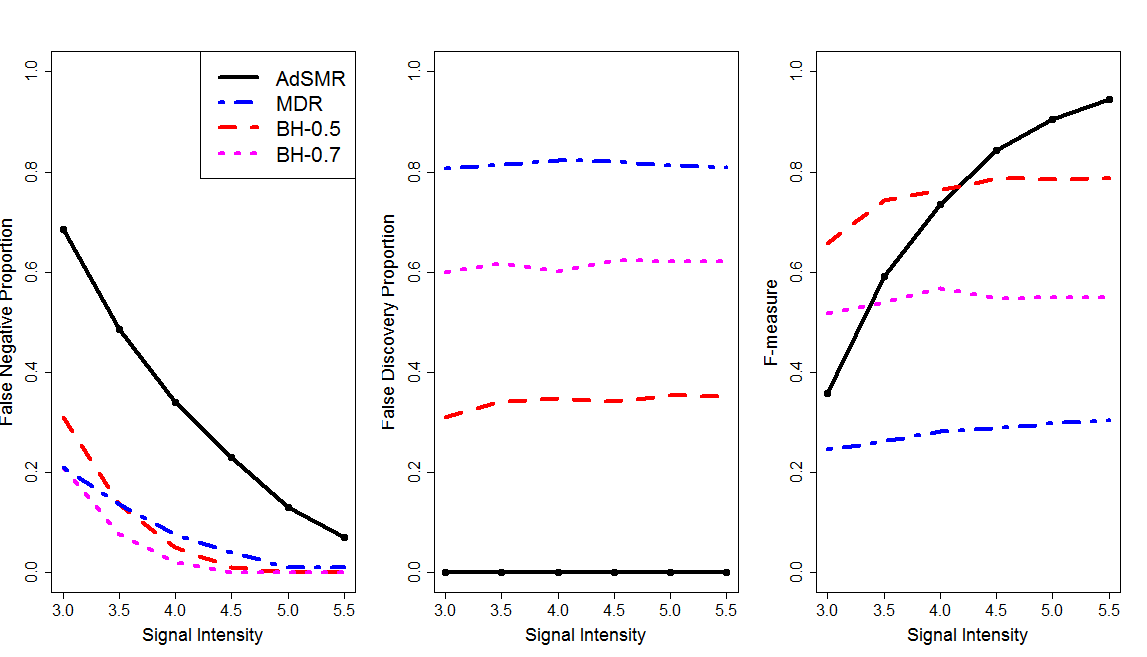}	
\end{figure}

Example 2 has block dependence with larger block size $l=200$. Figure \ref{Fig:largeL} compares all the methods under different signal sparsity and intensity levels. Compared to Example 1, the performance of AdSMR deteriorates a little for FN control, which agree with the theoretical insights in Theorem \ref{thm:unify} on the effect of block size.  
Overall, AdSMR still mostly outperforms other methods in FDP control and F-measure.

\begin{figure} [h!]
	\centering
	\caption{Comparison of methods in false negative proportion, false discovery proportion, and F-measure for sparse covariance matrix. The top row has $\pi = 0.1$, and the bottom row has $\pi = 0.02$.} \label{Fig:nonblock}
	\includegraphics[width=6in,height=2.2in]{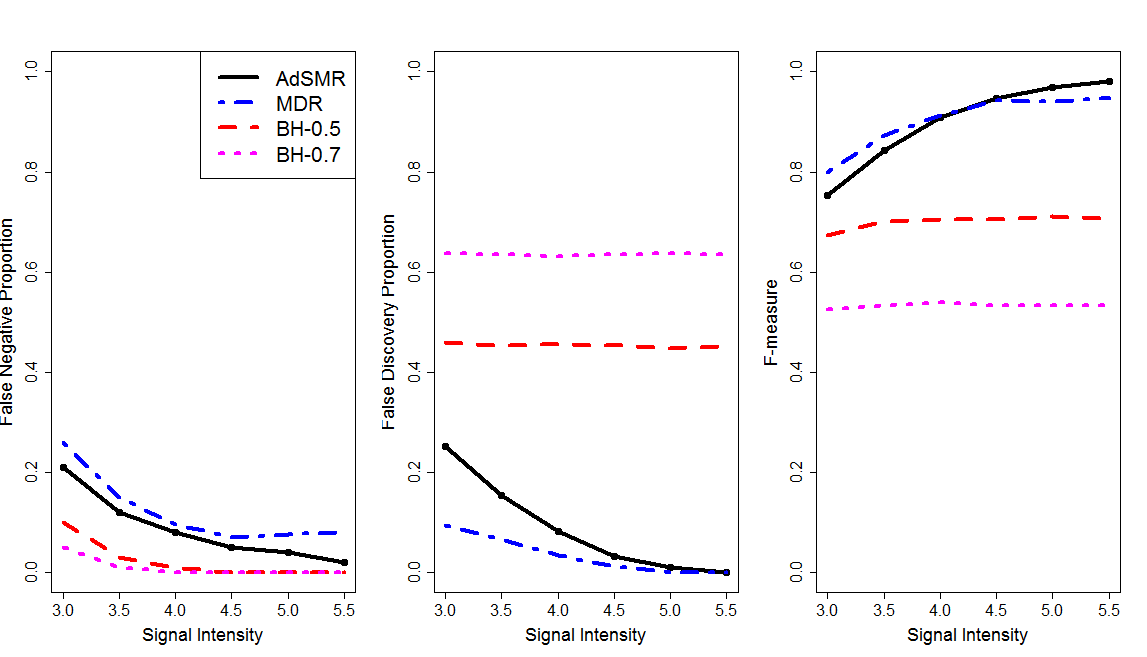}
	\includegraphics[width=6in,height=2.2in]{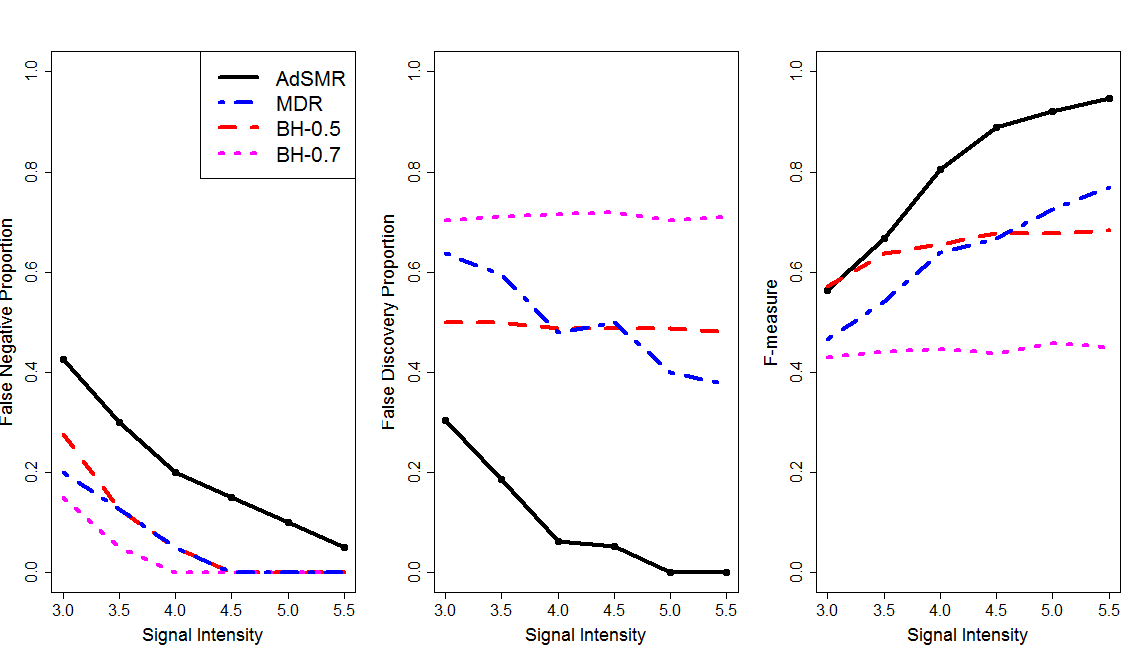}
\end{figure}

Example 3 simulates data with $m=1000$ and a sparse $\bf{\Sigma}$ whose nonzero elements are randomly located. The data generation process is similar to Model 3 in \cite{cai2013two}. Let $\Sigma^* = (\sigma_{ij})$, where $\sigma_{ii} = 1$, $\sigma_{ij} = 0.7 *$ Bernoulli$(1, 0.1)$ for $i<j$ and $\sigma_{ji} = \sigma_{ij}$. Then $\bf{\Sigma} = I^{1/2}(\Sigma^* + \delta I)/ (1+\delta) I^{1/2}$, where $\delta = |\lambda_{min} (\Sigma^*)| + 0.05$. 
Figure \ref{Fig:nonblock} shows that the performance of AdSMR is comparable to its performance in Example 1, although the covariance matrix in this example does not have a block structure. MDR performs relatively better in this example than in previous examples. AdSMR and MDR generally outperform high level BH-FDR procedures in this example.  

\begin{figure} [h!]
	\centering
	\caption{Comparison of methods in false negative proportion, false discovery proportion, and F-measure under dependence structure from a two-factor model. The top row has $\pi = 0.1$, and the bottom row has $\pi = 0.02$.} \label{Fig:factor}
	\includegraphics[width=6in,height=2.2in]{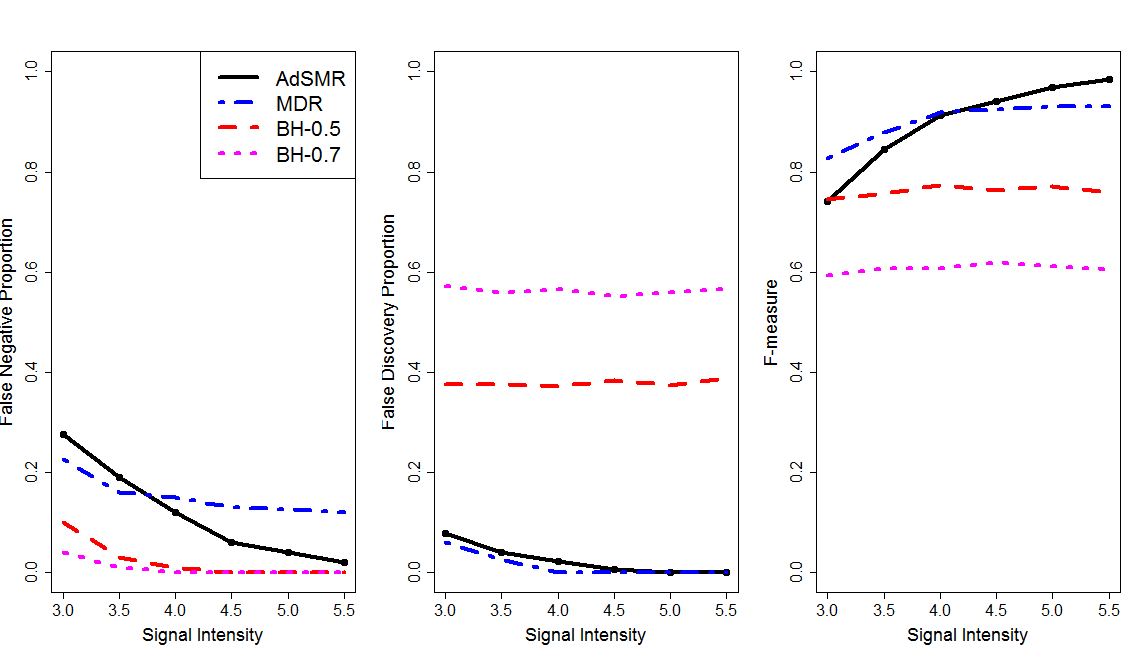}
	\includegraphics[width=6in,height=2.2in]{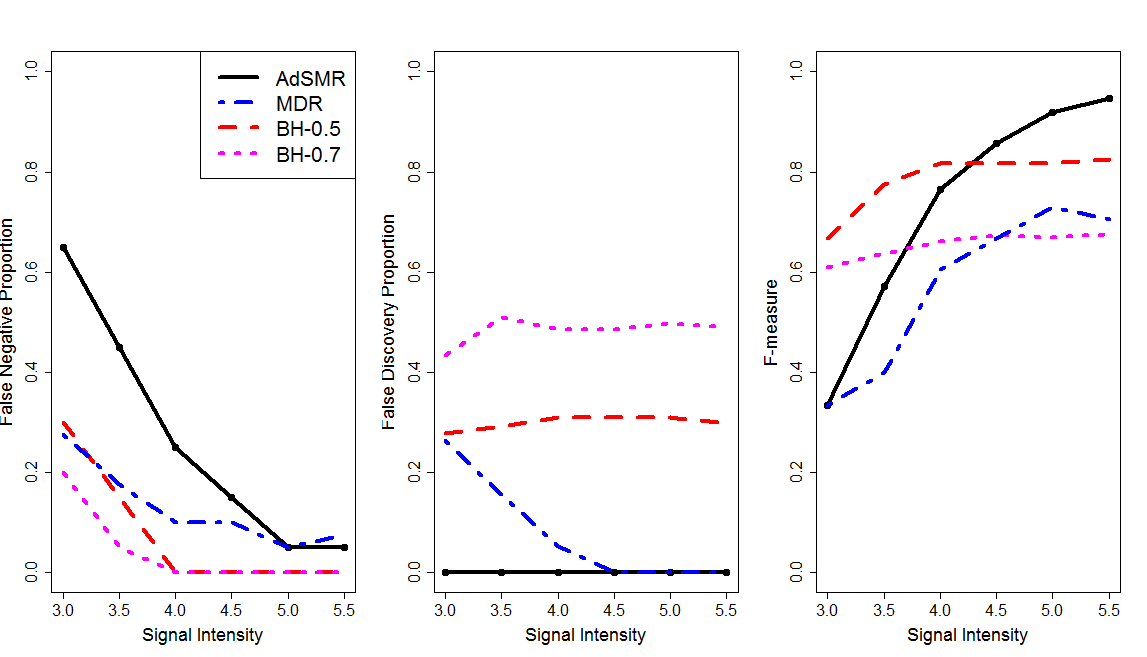}
\end{figure}

Example 4 considers dependence structure from a two-factor model as in \cite{fan2012estimating}. Let $\bf{\Sigma}$ be the correlation matrix of a random sample with size 100 of $m$-dimensional vector $\mathbf{X} = (X_{1}, \ldots, X_{m})$, where $X_j = \rho^{(1)}_{j} W^{(1)} + \rho^{(2)}_{j} W^{(2)} + H_{j}$, $W^{(1)}$ and $W^{(2)}$ are iid $N(0, 1)$, $\rho^{(1)}_{j}$ and $\rho^{(2)}_{j}$ are iid $U(-1, 1)$, and $H_j$ are iid $N(0, 1)$. Figure \ref{Fig:factor} shows that the performance of AdSMR is comparable to its performance in Example 2, where $\bf{\Sigma}$ has large diagonal blocks. MDR and high level BH-FDR procedures perform relatively better in this example than in the examples with block dependence.

In all the examples, the cut-off position of AdSMR automatically vary with signal sparsity and intensity and, as a result, when signal intensity becomes stronger and the noise and signal $p$-values are better separated, AdSMR controls both false negative and false positive better. 
BH-FDR with high nominal levels do not have such property as their FDPs remain high with increasing signal intensity. MDR method performs better than high level BH-FDR  in terms of adapting to signal intensity but worse than AdSMR when signals are more sparse.  Additional simulation for the empirical SMR of AdSMR is presented in Appendix \ref{sec:emp_SMR}. 
Finite-sample performance of the MR estimator along is demonstrated in Supplementary Material.

\section{Real Data Analysis} \label{sec:realdata}

Recent heritability analyses of GWAS data have suggested that a large proportion of variation of human height can be explained by all autosomal SNPs although only a small proportion of associated variants have been successfully identified. For example, \cite{yang2011gcta} showed that about $45\%$ of the variation can be accounted for by common SNPs from a sample of around $4000$ Australians with ancestry in the British Isles. In the analysis of \cite{kostem2013improving}, $62\%$ of the variation can be explained by all autosomal SNPs from the Northern Finland Birth Cohort of 1966 from $5319$ unrelated individuals. However, the discovered associated SNPs have only explained a small proportion of the total heritability. For example, in \cite{allen2010hundreds}, only $10.5\%$ of the variance can be explained by the $180$ SNPs with genome-wide significance. It indicates that there exist a large number of weak signals, for which current methods have limited power even with a sample size of  $10^3 \sim 10^4$. 

We obtained the GWAS height data from the Cohorte Lausannoise (CoLaus) study \citep{firmann2008colaus}. The dataset includes $1874$ subjects with available information on age, sex, and $340359$ autosomal SNPs with minor allele frequencies greater than $1\%$. We calculated the $p$-values corresponding to each SNP by fitting marginal linear models while adjusting  12 clinical covariates including sex, age, and the top 10 principal components.
The FDR method in \cite{benjamini1995controlling} cannot identify any associated SNPs based on the full set of SNPs due to small effects of individual SNPs. On the other hand, the Lasso procedure \citep{tibshirani1996regression} applied to the full dataset identifies only two candidate SNPs.

In order to reduce data dimension and carry as many true signals to subsequent analyses, we apply both AdSMR and MDR procedures. To implement the MR estimator to AdSMR, we first generate a set of $p$-value sequences under the global null using the permutation approach introduced in \cite{westfall1993resampling}.
In each permutation, the phenotype values are randomly shuffled and reassigned to individuals, the null $p$-values are calculated in each permutation with the correlation structure of the SNPs preserved. We replicate the permutation process $1000$ times. Based on the simulated null $p$-values, the MR estimate and
the cut-off position of AdSMR are calculated by the algorithm in Section \ref{sec:algorithm}. The estimated number of associated SNPs is equal to $743$; AdSMR selects $7,204$ SNPs, while MDR selects $110,778$ SNPs.

To examine the contribution of the selected SNPs, we perform heritability analysis using GCTA package (\cite{yang2011gcta}). First, we use GCTA to estimate the genetic relationship matrix (GRM) of all the GWAS SNPs and then fit a random effects model to estimate the proportion of variance explained by all the autosomal SNPs. 
We found that  $53.7\%$ of the phenotypic variance in height can be explained by all the autosomal SNPs. We then repeat the analysis on the partitioned SNP sets (e.g., the $7,204$ SNPs identified by AdSMR vs. the rest) to estimate the proportion of variance in height explained by the selected SNPs while adjusting for the remaining unselected SNPs. 
Note that the random effects model in GCTA is very different from the marginal linear regression model that we used to select SNPs. In addition, the heritability percentage is also different from the $R^2$ statistic of goodness-of-fit test for prediction.

Table \ref{tab:height} shows that AdSMR reduces the number of SNPs from $340,359$ to $7,204$, but the explained variability of the selected SNPs is still $53.7\%$, suggesting that almost all important SNPs to human height are retained by AdSMR for this dataset. MDR also explained $53.7\%$ of the height variation, but appears less efficient than AdSMR by selecting a much larger set of SNPs.

\begin{table}[h!] 
	\centering
	\caption{Numbers of selected SNPs and the estimated heritability.}
	\bgroup
	\def\arraystretch{1.2}
	\begin{tabular}{l c c }
		\hline
		& Number of SNPs & Estimated Heritability\\
		\hline
		Total & $340,359$  & $53.7\%$  \\
		AdSMR  &  $7,204$ & $53.7\%$ \\
		MDR & $110,778$ & $53.7\%$ \\
		\hline
	\end{tabular}
	\egroup
	\label{tab:height}
\end{table}

With the reduced set of SNPs from AdSMR, we can apply joint modeling using Lasso and further narrow down the number of SNPs to 1,563. By repeating the above GCTA analysis on the 1,563 selected SNPs vs. the rest, the estimated heritability
is still $53.7\%$. These selected SNPs provide promising candidates for further downstream analyses such as gene annotation, pathway mapping, polygenic risk score, etc. 





\vspace{-0.1in}
\section{Conclusion and Further Discussion}

In this paper, we consider the problem of efficient signal inclusion under dependence. Our motivation comes from Big Data applications where sample sizes are relatively limited with respect to data dimension, and there are great needs to significantly reduce data dimension while retaining as many true signals as possible for subsequent analyses. 
However, in applications where signals are much rarer than noise, inference for false negative control based on signal information requires different techniques from those used for false positive control. Furthermore, the dependence among variables, especially between noise and signal variables, adds another layer of difficulty.   
To address these challenges, we develop data-adaptive methods whose implementations do not need the signal information. 
Nevertheless, the cut-off position of the proposed AdSMR procedure automatically vary with signal sparsity and intensity and, as a result, a high proportion of signals can be selected without incurring many unnecessary false positives. 
When signal intensity becomes stronger and the noise and signal $p$-values are better separated, the cut-off position of AdSMR controls both false negative and false positive better. These properties are presented in Theorem \ref{thm:control_2} - \ref{thm:unify} and Corollary \ref{cor:FP} - \ref{cor:typeI}, and illustrated in simulation examples where signal intensity increases in each setting.

We have also proposed a new measure, SMR, for false negative control. The notion of SMR includes two parameters: the FN proportion $\epsilon$ and the SMR control level $\alpha$. The methods developed in the paper, cvSMR and AdSMR, control SMR at different levels.
The construction of cvSMR is more in line with the existing techniques based on marginal distribution of $p$-values, whereas AdSMR benefits by new explorations on the concentration properties of the ordered $p$-values under dependence. 
Both methods control SMR for arbitrarily small $\epsilon$. However, AdSMR is shown to incur much less false positives.

We note that false positives can be further reduced by allowing a small fixed FN proportion. Such extension on AdSMR would involve more delicate analyses on the concentration properties of order statistics under dependence. The estimator $\hat s$ implemented in cvSMR and AdSMR is another subject to be re-investigated under the new request. Detailed studies are deferred to future research.  
Another interesting topic for future research is SMR control for specific data generating models. While the current paper considers a general $p$-value model without model assumptions on test statistics, it will be interesting to relate conditions in the paper to parameters of the specific models. We expect that the characterizations would be very different for models with sparse or dense signal component.

\vspace{-0.1in}
\section*{Acknowledgment}
The authors are grateful to Dr. Peter Vollenweider and Dr. Gerard Waeber, PIs of the CoLaus study, and Dr. Meg Ehm and Dr. Matthew Nelson, collaborators at GlaxoSmithKline, for providing the CoLaus phenotype and genetic data. The authors appreciate the very helpful comments and suggestions from an associate editor and three reviewers.  Dr. Jeng was partially supported by National Human Genome Research Institute of the National Institute of Health under grant R03HG008642. Dr. Tzeng was partially supported by National Institutes of Health grant P01CA142538.

\vspace{-0.1in}
\section*{Appendix}

Appendix \ref{sec:proof_prop} - \ref{sec:proof_index} provide proofs of Proposition \ref{prop:control_1} and Theorem \ref{thm:control_2}. 
A toy example demonstrating the AdSMR algorithm is shown in  Appendix \ref{sec:example}, and additional simulation results are presented in Appendix \ref{sec:emp_SMR}. Proofs of Theorem \ref{thm:FP}, Theorem \ref{thm:unify}, and Corollary \ref{cor:typeI} and more simulation results are provided in Supplementary Material. 

\begin{small}

\vspace{-0.1in}
\subsection{Proof of Proposition \ref{prop:control_1}} \label{sec:proof_prop}

For notation simplicity, let $j^* = \min\{j\geq 1: p_{(\hat s +j)} \leq \alpha_j\}$, where $\alpha_j = (j/m)\alpha$. Then ${k}^*_{cv} = \hat s + j^*$. 
Denote $TP(k)$ and $FP(k)$ as the numbers of true positives and false positives in the top $\{1, \ldots, k\}$ candidates, then ${k}^*_{cv} = TP({k}^*_{cv}) + FP({k}^*_{cv})$. We also have $s = TP({k}^*_{cv}) + FN({k}^*_{cv})$. Now, for any $\epsilon>0$,
\vspace{-0.2in}
\begin{eqnarray} \label{mr}
SMR^\epsilon({k}^*_{cv}) & = & P(FN({k}^*_{cv})/ s > \epsilon )  = P(TP({k}^*_{cv}) < s- \epsilon s ) = P(FP({k}^*_{cv}) > {k}^*_{cv} - (1-\epsilon)s) \nonumber \\
&\leq& P(FP({k}^*_{cv}) > \hat s + j^* - (1-\epsilon) s, ~\hat s\geq (1-\epsilon) s) + P(\hat s<(1-\epsilon)s) \nonumber \\
& \le &  P(FP({k}^*_{cv}) > j^*) + o(1),
\end{eqnarray}
where the last step is by the consistency of $\hat s$. 

In the case of $\hat s \leq t_1$, we have ${k}^*_{cv} = \hat s \le t_1$ and $j^*=0$, then
\begin{eqnarray} \label{3}
P(FP({k}^*_{cv}) > j^*) & \leq & P(FP(t_1)>0)  \nonumber \\
& = & P(\text{at least one of the } m_0 \text{ noise variables rank ahead of } t_1)  \nonumber \\
& \le & P(\text{at least one of the } m \text{ noise variables have p-value } < \alpha_m/m)  \nonumber \\
& \le & m {\alpha_m \over m} = \alpha_m = o(1). 
\end{eqnarray}
Combining (\ref{mr}) and (\ref{3}) implies (\ref{eq:control_1}). 

In the case of $\hat s > t_1$, consider the conditional probability $P(FP({k}^*_{cv})>j^* | P_1^1, \ldots, P_s^1)$. By Markov's inequality, 
\begin{eqnarray*} \label{4}
P(FP({k}^*_{cv})>j^* | P_1^1, \ldots, P_s^1) & \le & {1\over j^*}E(FP({k}^*_{cv}) | P_1^1, \ldots, P_s^1)   =  {1\over j^*} E(\sum_{j=1}^{m_0} 1(P_j^0 \le \alpha_{j^*}| P_1^1, \ldots, P_s^1) \nonumber \\
& = & {1\over j^*} \sum_{j=1}^{m_0} P(P_j^0 \le \alpha_{j^*} | P_1^1, \ldots, P_s^1) \le  {m_0\over j^*} \alpha_{j^*}  =  {m_0\over j^*} {j^* \over m} \alpha ~ \le ~ \alpha,
\end{eqnarray*}
where the fourth step is by condition (\ref{cond:jointP_1}). The above implies 
\begin{equation} \label{4}
P(FP({k}^*_{cv})>j^*) \le \alpha
\end{equation}
Combining (\ref{mr}) and (\ref{4}) gives (\ref{eq:control_1}).

\vspace{-0.1in}
\subsection{Proof of Theorem \ref{thm:control_2}} \label{sec:proof_adSMR}

We consider $\bar l \le \epsilon s/2$ and $\bar l > \epsilon s/2$ separately. 

When $\bar l \le \epsilon s/2$, the proof is relatively straight-forward. First, by the definitions of $T_2$ and $\bar l$, 
\[
s = TP(T_2) = T_2 - FP(T_2) > T_2-\bar l \ge T_2-\epsilon s/2, 
\]
which implies $T_2 < (1 + \epsilon/2) s$ with probability tending to 1. On the other hand, 
\[
k^* \ge \hat s > (1-\epsilon/4) s > {1-\epsilon/4 \over 1+ \epsilon/2} T_2 > T_2 - (2\epsilon /3)T_2,
\]
which implies that 
\[
TP(k^*) \ge TP(T_2) - (2\epsilon /3)T_2 \ge s - (2\epsilon /3) (1 + \epsilon/2) s > (1-\epsilon) s
\]
with probability tending to 1. This conclude the case with $\bar l \le \epsilon s/2$. 

The following proof is for $\bar l > \epsilon s/2$. 
By the definition of $SMR^\epsilon$, it is enough to show  that for any $\epsilon>0$, 
\begin{equation} \label{eq:0}
P(FN(k^*) > \epsilon s) \to 0. 
\end{equation}

The case $\hat s \le t_1$ can be proved by similar arguments leading to (\ref{3}).

Consider the case $\hat{s}> t_1$. Without loss of generality, assume $\epsilon s$ is an integer. Denote $T_2^{(1-\epsilon)s}$ as the position for the $(1-\epsilon) s$-th signal. Then
\[
\{FN(k^*) > \epsilon s\} \subseteq \{k^*<T_2^{(1-\epsilon) s} \} \subseteq \{\exists j \in \{1, \ldots, T_2^{(1-\epsilon) s}-\hat s\} ~ s.t. P_{(\hat s+j)} \le b_j \}
\]
Note that $P_{(1)}, \ldots, P_{(T_2)}$ are composed of $P_{(1)}^1, \ldots, P_{(s)}^1$ and $P_{(1)}^0, \ldots, P_{(L)}^0$. Denote $b_{j_r} $  and $b_{j_q} $ as the critical values corresponding to $P_{(r)}^0$ and $P_{(q)}^1$ in  $\{P_{(\hat s+1)}, \ldots, P_{(T_2^{(1-\epsilon) s})}\}$, respectively. Then
\begin{eqnarray} \label{1.1}
P(FN(k^*) > \epsilon s) & \le & P(\exists j \in \{1, \ldots, T_2^{(1-\epsilon) s}-\hat s\} ~ s.t. P_{(\hat s+j)} \le b_j ) \nonumber\\
& \le & P(\exists r ~ s.t. P^0_{(r)} \le b_{j_r}) + P(\exists q ~ s.t. P^1_{(q)} \le b_{j_q})
\end{eqnarray}
Let $P^0_{(r_q)}$ be the largest $P^0_{(r)}$ before $P^1_{(q)}$. The following lemma shows the relationships between $r$ and $j_r$ and between $r_q$ and $j_q$. 
\begin{lemma} \label{lemma:index}
	Given a consistent estimator $\hat s$ for the  number of signals and  $\bar l > \epsilon s/2$, we have
	\begin{equation} \label{eq:1}
	j_r + \epsilon s/2 < r < L \le \bar{l}
	\end{equation}
	and
	\vspace{-0.2in}
	\begin{equation} \label{eq:2}
	j_q + \epsilon s/2 < r_q < L \le \bar{l}
	\end{equation}
	with high probability.
	\end{lemma}
	
	Now consider the first term of (\ref{1.1}). By condition (\ref{cond:jointP_2}), 
	\begin{equation*}
	P(P^0_{(r)} \le b_{j_r}  | P_{1}^1, \ldots, P_{s}^1) \le  P(B_{r} \le b_{j_r} ),
	\end{equation*}
	where $B_{r}$ is a random variable following $Beta(\nu_1, \nu_2)$ with $\nu_1=r$ and $\nu_2 = m-s-r+1$. Further, by (\ref{eq:1}),
	\begin{equation*}
	P(B_r \le b_{j_r}  ) < P(B_r \le b_{r- \epsilon s/2}  )  \le   P(B_{\bar{l}} \le b_{\bar{l}- \epsilon s/2}  ),
	\end{equation*}
	where the second inequality is by the properties of Beta distribution that $B_{(r)}$ is less positively skewed as $r$ increases and the change of skewness gets slower as $r$ approaches to $(m-s)/2$. Then the above implies 
	\begin{equation} \label{1.2}
	P(P^0_{(r)} \le b_{j_r}) \le P(B_{\bar{l}} \le b_{\bar{l}- \epsilon s/2}).
	\end{equation} 
	
	Consider the second term of (\ref{1.1}). Clearly $P(P^1_{(q)} \le b_{j_q}) < P(P^0_{(r_q)}  \le  b_{j_q})$. Similar arguments as above combining condition (\ref{cond:jointP_2}) and (\ref{eq:2}) in Lemma \ref{lemma:index} give 	
	\[
	P(P^0_{(r_q)}  \le  b_{j_q}  | P_{(1)}^1, \ldots, P_{(s)}^1) \le P(B_{r_q} \le b_{j_q} ) 
	\le  P(B_{r_q} \le b_{r_q- \epsilon s/2} ) 
	\le  P(B_{\bar{l}} \le b_{\bar{l}- \epsilon s/2} ),
	\]
	which implies 
	\begin{equation} \label{1.3}
		P(P^1_{(q)} \le b_{j_q}) < P(P^0_{(r_q)}  \le  b_{j_q})  \le P(B_{\bar{l}} \le b_{\bar{l}- \epsilon s/2}).
	\end{equation} 
	
	Combining (\ref{1.2}) and (\ref{1.3}) with (\ref{1.1}) gives
	\begin{eqnarray} \label{1.6} 
	P(FN(k^*) > \epsilon s) & \le & (\bar{l} + s) \cdot  P(B_{\bar{l}} \le b_{\bar{l}- \epsilon s/2} ).
	\end{eqnarray}

	Note that there is no explicit form for the cumulative distribution function of Beta distribution. To derive the probability in (\ref{1.6}), let 
	\[
	F = {\nu_2 B_{\bar{l}}  \over \nu_1(1-B_{\bar{l}}) }.
	\]
	By the relationship between Beta and F distributions, $F$ has an $F_{2 \nu_1, 2 \nu_2}$ distribution \citep{Johnson1970}. In our case, $1 \ll \nu_1 \ll \nu_2$ by condition (\ref{cond:barL}), then $F_{2 \nu_1, 2 \nu_2}$ is highly concentrated at $1$ with mean $\approx 1$ and variance $\approx 1/\nu_1$.
	On the other hand,  we know that the median of Beta distribution is bounded by its mean for $1<\nu_1 <\nu_2$, then 
	\[
	b_{\bar{l}- \epsilon s/2} < E(B_{\bar{l}- \epsilon s/2}) = {\bar{l}- \epsilon s/2 \over m-\hat s +1} 
	\]
	and
	\vspace{-0.2in}
	\begin{eqnarray} \label{1.7}
	P(B_{\bar{l}} \le b_{\bar{l}- \epsilon s/2} ) &=& P( F \le {\nu_2 \over \nu_1} {b_{\bar{l}- \epsilon s/2} \over 1-b_{\bar{l}- \epsilon s/2}} ) \le  P(F \le  {\nu_2 \over \nu_1} b_{\bar{l}-\epsilon s/2} + C {\nu_2 \over \nu_1} b^2_{\bar{l}-\epsilon s/2})  \nonumber \\
	&\le& P(F \le  {\nu_2 \over \nu_1} {\bar{l}- \epsilon s/2 \over m-\hat s +1} + C{\bar{l} \over m}) \le  P(F \le  1- C{\epsilon s \over \bar{l}}),
	\end{eqnarray}
	where the last step is by condition (\ref{cond:barL}). 
	Further, let 
	\[
	Z = {F^{1/3}(1-{2\over 9\nu_2}) + {2 \over 9\nu_1} -1 
		\over \sqrt{{2F^{2/3} \over 9 \nu_2} + {2 \over 9\nu_1}}},
		\]
		then by Wilson-Hilferty approximation to $\chi^2$ and the fact that $F$ is a ratio of $\chi^2$ distribution, $Z$ approximately follows $N(0,1)$ distribution \citep{Johnson1970}. It is clear that 
		\[
		Z < {F^{1/3} + {2 \over 9\nu_1} -1 
			\over \sqrt{{2 \over 9\nu_1}}},
			\]
			then 
			\begin{eqnarray} \label{1.8}
			P(F \le  1-C{\epsilon s \over \bar{l}}) &<& P(\left(\sqrt{{2 \over 9\nu_1}}Z +1 - {2 \over 9\nu_1}\right)^3 \le  1-C {\epsilon s \over \bar{l}}) \nonumber \\
			&=& P(Z \le \sqrt{{9\nu_1 \over 2}} \left((1-C{\epsilon s \over \bar{l}})^{1/3} -1+{2 \over 9\nu_1}\right))
			\end{eqnarray}
			Apply Taylor expansion on $(1-C{\epsilon s \over \bar{l}})^{1/3}$, 
			\[
			(1-C{\epsilon s \over \bar{l}})^{1/3} - 1 +{2 \over 9\nu_1} \le -C {\epsilon s \over \bar{l}}  +{2 \over 9\bar{l}} \le -C {\epsilon s \over \bar{l}}
			\]
			as $s \gg 1$ is implied by condition (\ref{cond:barL}). Then
			\begin{eqnarray} \label{1.9}
			P(Z \le \sqrt{{9\nu_1 \over 2}} \left((1-C{\epsilon s \over \bar{l}})^{1/3} -1+{2 \over 9\nu_1}\right))& \le &  P\left(Z \le \sqrt{{9\bar{l} \over 2}}(-C{\epsilon s \over \bar{l}}) \right) \nonumber \\ 
			& \le & P\left(Z \le - C {s \over \sqrt{\bar{l}}} \right) 
			\end{eqnarray}
			Finally, by the Normal approximation of $Z$ and Mill's inequality, 
			\begin{equation} \label{1.10}
			(\bar{l} + s) P\left(Z \le - C {s \over \sqrt{\bar{l}}} \right) \le C \bar{l}^{3/2} \exp(-{Cs^2 \over \bar{l}}) + C s \bar{l}^{1/2} \exp(-{Cs^2 \over \bar{l}}) = o(1),
			\end{equation}
			where the last step is by condition (\ref{cond:barL}). Summarizing (\ref{1.6}) - (\ref{1.10}) gives (\ref{eq:0}). 

\vspace{-0.15in}
\subsection{Proof of Lemma \ref{lemma:index}} \label{sec:proof_index}

Consider (\ref{eq:1}) first. At the position of $\hat s + j_r$, 
\[
\hat s + j_r = FP(\hat s + j_r) + TP(\hat s + j_r) \le r + TP(T_2^{(1-\epsilon) s})  = r+ (1-\epsilon)s.   
\]
On the other hand, the consistency of $\hat s$ implies $\hat s > (1-\epsilon/2) s$ with high probability, then
\[
\hat s + j_r > (1-\epsilon/2) s + j_r
\]
with high probability. Combining the above gives 
\[
r > j_r + \epsilon s/2
\]
with high probability. Further, by the definitions of $r$, $L$ and $\bar{l}$, 
\[
r \le FP(T_2^{(1-\epsilon)s}) \le FP(T_2) = L \le \bar{l}.
\]
Then (\ref{eq:1}) follows. 

(\ref{eq:2}) can be proved in the similar way as above given the fact that $FP(\hat s + j_q) = r_q$.

\vspace{-0.1in}
\subsection{A Toy example for AdSMR algorithm} \label{sec:example}

We provide a simple example demonstrating the AdSMR algorithm presented at the end of Section \ref{sec:algorithm}. 
Suppose the data yield $\hat s = 1$ and $10$ ordered $p$-values 
\[
\{0.02, 0.11, 0.12, 0.21, 0.36, 0.49, 0.69, 0.77, 0.87, 0.99\}.
\] 
Recall the cut-off position in (\ref{cut_weak}) 
\[
{k}^*_{cv} = \hat s + \min \{j \ge 1: p_{(\hat s+j)}\leq b_j\}1\{\hat{s}>t_1\},
\]
where $t_1=\max\{j: p_{(j)} < \alpha_m/m\}$ with $\alpha_m=1/\sqrt{\log m}$.  For this example with $m=10$,  $t_1=\max\{j: p_{(j)} < 0.066\} = 1$. Then, the indicator function  $1\{\hat{s}>t_1\} = 0$; and ${k}^* = \hat s = 1$. 

On the other hand, if the data yield  $\hat{s} = 2$, then the indicator function  $1\{\hat{s}>t_1\} = 1$,  and the procedure continues to check $p_{(2+j)} \le b_j$. In this example, $p_{(2+j)} = 0.12, 0.21, 0.36 \ldots$ and $b_j = 1/8, 2/8, 3/8, \ldots$ for $j= 1, 2, 3, \ldots$. Then $k^* = \hat s + 1 = 3$.

\subsection{Empirical SMR of AdSMR}\label{sec:emp_SMR}

We illustrate the empirical SMR of AdSMR with the MR estimator under the settings of Example 1 in simulation. Recall that Theorem \ref{thm:unify} has  $SMR^{\epsilon}(k^*) \to 0$ for any constant $\epsilon>0$ under suitable conditions. Table \ref{tab:empirical_SMR} reports the empirical $SMR^\epsilon$ of AdSMR for $\epsilon = 0.1, 0.2$ and $0.3$ from 100 replications. The results agree with Figure \ref{Fig:smallL}. For example, Figure \ref{Fig:smallL} shows the median of FN proportion around 0.18 for AdSMR when $\pi=0.02$ and $\mu=4.5$. Table \ref{tab:empirical_SMR} shows that when $\mu=4.5$, the FN proportion of AdSMR is greater than 0.1 in 65 (out of 100) replications, greater than 0.2 in 29 replications, and greater than 0.3 in 0 replications. The empirical SMR decreases as $\mu$ increases.

\begin{table}[!h] 
	\small
	\centering
	\caption{Empirical $SMR^{\epsilon}(k^*)$ of AdSMR in the setting of Figure \ref{Fig:smallL} with $\pi = 0.02$.} \label{tab:empirical_SMR}
	\begin{tabular}{|l|c c c c c|}
		\hline
		&  \multicolumn{5}{c|}{$\mu$} \\
		&  $4.5$ & $5$ & $5.5$ & $6$ & $6.5$ \\
		\hline
		$\epsilon = 0.1$ & 0.65 & 0.43 & 0.03 & 0 & 0\\
		$\epsilon = 0.2$ & 0.29 & 0 & 0 & 0 & 0\\
		$\epsilon = 0.3$ & 0 & 0 & 0 & 0 & 0 \\
		\hline
	\end{tabular}
\end{table}

\end{small}

\bibliographystyle{Chicago}
\bibliography{myresearch}

\begin{thebibliography}{}

\bibitem[\protect\citeauthoryear{Abraham, Kowalczyk, Zobel, and Inouye}{Abraham
  et~al.}{2013}]{abraham2013}
Abraham, G., A.~Kowalczyk, J.~Zobel, and M.~Inouye (2013).
\newblock Performance and robustness of penalized and unpenalized methods for
  genetic prediction of complex human disease.
\newblock {\em Genetic Epidemiology\/}~{\em 37\/}(2), 184--195.

\bibitem[\protect\citeauthoryear{Allen, Estrada, Lettre, Berndt, Weedon,
  Rivadeneira, Willer, Jackson, Vedantam, Raychaudhuri, et~al.}{Allen
  et~al.}{2010}]{allen2010hundreds}
Allen, H.~L., K.~Estrada, G.~Lettre, S.~I. Berndt, M.~N. Weedon,
  F.~Rivadeneira, C.~J. Willer, A.~U. Jackson, S.~Vedantam, S.~Raychaudhuri,
  et~al. (2010).
\newblock Hundreds of variants clustered in genomic loci and biological
  pathways affect human height.
\newblock {\em Nature\/}~{\em 467\/}(7317), 832--838.

\bibitem[\protect\citeauthoryear{Benjamini and Hochberg}{Benjamini and
  Hochberg}{1995}]{benjamini1995controlling}
Benjamini, Y. and Y.~Hochberg (1995).
\newblock Controlling the false discovery rate: a practical and powerful
  approach to multiple testing.
\newblock {\em Journal of the Royal Statistical Society. Series B\/}, 289--300.

\bibitem[\protect\citeauthoryear{Cai, Liu, and Xia}{Cai
  et~al.}{2013}]{cai2013two}
Cai, T., W.~Liu, and Y.~Xia (2013).
\newblock Two-sample covariance matrix testing and support recovery in
  high-dimensional and sparse settings.
\newblock {\em Journal of the American Statistical Association\/}~{\em
  108\/}(501), 265--277.

\bibitem[\protect\citeauthoryear{Cai, Jin, Low, et~al.}{Cai
  et~al.}{2007}]{cai2007estimation}
Cai, T.~T., J.~Jin, M.~G. Low, et~al. (2007).
\newblock Estimation and confidence sets for sparse normal mixtures.
\newblock {\em The Annals of Statistics\/}~{\em 35\/}(6), 2421--2449.

\bibitem[\protect\citeauthoryear{Cai and Sun}{Cai and
  Sun}{2017}]{cai2016optimal}
Cai, T.~T. and W.~Sun (2017).
\newblock Optimal screening and discovery of sparse signals with applications
  to multistage high-throughput studies.
\newblock {\em Journal of the Royal Statistical Society: Series B\/}~{\em
  79\/}(1), 197--223.

\bibitem[\protect\citeauthoryear{Efron}{Efron}{2007}]{Efron07}
Efron, B. (2007).
\newblock Correlation and large-scale simultaneous significance testing.
\newblock {\em Journal of the American Statistical Association\/}~{\em 102},
  93--103.

\bibitem[\protect\citeauthoryear{Fan, Han, and Gu}{Fan
  et~al.}{2012}]{fan2012estimating}
Fan, J., X.~Han, and W.~Gu (2012).
\newblock Estimating false discovery proportion under arbitrary covariance
  dependence.
\newblock {\em Journal of the American Statistical Association\/}~{\em
  107\/}(499), 1019--1035.

\bibitem[\protect\citeauthoryear{Fan and Lv}{Fan and Lv}{2008}]{fan2008sure}
Fan, J. and J.~Lv (2008).
\newblock Sure independence screening for ultrahigh dimensional feature space.
\newblock {\em Journal of the Royal Statistical Society: Series B (Statistical
  Methodology)\/}~{\em 70\/}(5), 849--911.

\bibitem[\protect\citeauthoryear{Firmann, Mayor, Vidal, Bochud, P{\'e}coud,
  Hayoz, Paccaud, Preisig, Song, Yuan, et~al.}{Firmann
  et~al.}{2008}]{firmann2008colaus}
Firmann, M., V.~Mayor, P.~M. Vidal, M.~Bochud, A.~P{\'e}coud, D.~Hayoz,
  F.~Paccaud, M.~Preisig, K.~S. Song, X.~Yuan, et~al. (2008).
\newblock The colaus study: a population-based study to investigate the
  epidemiology and genetic determinants of cardiovascular risk factors and
  metabolic syndrome.
\newblock {\em BMC cardiovascular disorders\/}~{\em 8\/}(1), 6.

\bibitem[\protect\citeauthoryear{Genovese and Wasserman}{Genovese and
  Wasserman}{2004}]{genovese2004stochastic}
Genovese, C. and L.~Wasserman (2004).
\newblock A stochastic process approach to false discovery control.
\newblock {\em The Annals of Statistics\/}, 1035--1061.

\bibitem[\protect\citeauthoryear{Hung, Lin, Chen, Wang, Huang, and Tzeng}{Hung
  et~al.}{2016}]{Hung2016}
Hung, H., Y.~Lin, P.~Chen, C.~Wang, S.~Huang, and J.~Tzeng (2016).
\newblock Detection of gene-gene interactions using multistage sparse and
  low-rank regression.
\newblock {\em Biometrics\/}~{\em 72}, 85–94.

\bibitem[\protect\citeauthoryear{Jin and Cai}{Jin and
  Cai}{2007}]{jin2007estimating}
Jin, J. and T.~T. Cai (2007).
\newblock Estimating the null and the proportion of nonnull effects in
  large-scale multiple comparisons.
\newblock {\em Journal of the American Statistical Association\/}~{\em
  102\/}(478), 495--506.

\bibitem[\protect\citeauthoryear{Johnson and Kotz}{Johnson and
  Kotz}{1970}]{Johnson1970}
Johnson, N.~L. and S.~Kotz (1970).
\newblock {\em Continuous Univariate Distributions-2.}, Volume~2.
\newblock John Wiley \& Sons.

\bibitem[\protect\citeauthoryear{Kao, Leung, L.W.C., Yip, and Yap}{Kao
  et~al.}{2017}]{Kao2017}
Kao, P., K.~H. Leung, C.~L.W.C., S.~Yip, and M.~Yap (2017).
\newblock Pathway analysis of complex diseases for gwas, extending to consider
  rare variants, multi-omics and interactions.
\newblock {\em Biochim Biophys Acta\/}~{\em 1861}, 335–353.

\bibitem[\protect\citeauthoryear{Kostem and Eskin}{Kostem and
  Eskin}{2013}]{kostem2013improving}
Kostem, E. and E.~Eskin (2013).
\newblock Improving the accuracy and efficiency of partitioning heritability
  into the contributions of genomic regions.
\newblock {\em The American Journal of Human Genetics\/}~{\em 92\/}(4),
  558--564.

\bibitem[\protect\citeauthoryear{Lehmann and Romano}{Lehmann and
  Romano}{2005}]{Lehmann2005}
Lehmann, E.~L. and J.~P. Romano (2005).
\newblock Generalizations of the familywise error rate.
\newblock {\em The Annals of Statistics\/}~{\em 33\/}(3), 1138--54.

\bibitem[\protect\citeauthoryear{Meinshausen and Rice}{Meinshausen and
  Rice}{2006}]{meinshausen2006estimating}
Meinshausen, N. and J.~Rice (2006).
\newblock Estimating the proportion of false null hypotheses among a large
  number of independently tested hypotheses.
\newblock {\em The Annals of Statistics\/}~{\em 34\/}(1), 373--393.

\bibitem[\protect\citeauthoryear{Powers}{Powers}{2011}]{F-measure2011}
Powers, D. (2011).
\newblock Evaluation: From precision, recall and f-measure to roc,
  informedness, markedness and correlation.
\newblock {\em J. of Machine Learning Technologies\/}~{\em 2\/}(1), 37--63.

\bibitem[\protect\citeauthoryear{Sarkar}{Sarkar}{2006}]{Sarkar2006}
Sarkar, S.~K. (2006).
\newblock False discovery and false nondiscovery rates in single-step multiple
  testing procedures.
\newblock {\em The Annals of Statistics\/}~{\em 34}, 394--415.

\bibitem[\protect\citeauthoryear{Storey}{Storey}{2002}]{storey2002direct}
Storey, J.~D. (2002).
\newblock A direct approach to false discovery rates.
\newblock {\em Journal of the Royal Statistical Society: Series B\/}~{\em
  64\/}(3), 479--498.

\bibitem[\protect\citeauthoryear{Tibshirani}{Tibshirani}{1996}]{tibshirani1996regression}
Tibshirani, R. (1996).
\newblock Regression shrinkage and selection via the lasso.
\newblock {\em Journal of the Royal Statistical Society. Series B
  (Methodological)\/}, 267--288.

\bibitem[\protect\citeauthoryear{Waldmann, M{\'e}sz{\'a}ros, Gredler, Fuerst,
  and S{\"o}lkner}{Waldmann et~al.}{2013}]{waldmann2013}
Waldmann, P., G.~M{\'e}sz{\'a}ros, B.~Gredler, C.~Fuerst, and J.~S{\"o}lkner
  (2013).
\newblock Evaluation of the lasso and the elastic net in genome-wide
  association studies.
\newblock {\em Frontiers in genetics\/}~{\em 4}, 270.

\bibitem[\protect\citeauthoryear{Westfall and Young}{Westfall and
  Young}{1993}]{westfall1993resampling}
Westfall, P.~H. and S.~S. Young (1993).
\newblock {\em Resampling-based multiple testing: Examples and methods for
  p-value adjustment}, Volume 279.
\newblock John Wiley \& Sons.

\bibitem[\protect\citeauthoryear{Wu, Devlin, Ringquist, Trucco, and Roeder}{Wu
  et~al.}{2010}]{wu2010}
Wu, J., B.~Devlin, S.~Ringquist, M.~Trucco, and K.~Roeder (2010).
\newblock Screen and clean: a tool for identifying interactions in genome-wide
  association studies.
\newblock {\em Genetic epidemiology\/}~{\em 34\/}(3), 275--285.

\bibitem[\protect\citeauthoryear{Wu, Chen, Hastie, Sobel, and Lange}{Wu
  et~al.}{2009}]{wu2009}
Wu, T.~T., Y.~F. Chen, T.~Hastie, E.~Sobel, and K.~Lange (2009).
\newblock Genome-wide association analysis by lasso penalized logistic
  regression.
\newblock {\em Bioinformatics\/}~{\em 25\/}(6), 714--721.

\bibitem[\protect\citeauthoryear{Yang, Lee, Goddard, and Visscher}{Yang
  et~al.}{2011}]{yang2011gcta}
Yang, J., S.~H. Lee, M.~E. Goddard, and P.~M. Visscher (2011).
\newblock Gcta: a tool for genome-wide complex trait analysis.
\newblock {\em The American Journal of Human Genetics\/}~{\em 88\/}(1), 76--82.

\bibitem[\protect\citeauthoryear{Zhou, Alexander, Sehl, Sinsheimer, Sobel, and
  Lange}{Zhou et~al.}{2011}]{zhou2011}
Zhou, H., D.~H. Alexander, M.~E. Sehl, J.~S. Sinsheimer, E.~Sobel, and K.~Lange
  (2011).
\newblock Penalized regression for genome-wide association screening of
  sequence data.
\newblock In {\em Biocomputing 2011}, pp.\  106--117. World Scientific.

\end{thebibliography}

\end{document}